\RequirePackage{etex} % Ensure extended capacity is loaded early
\documentclass[conference]{IEEEtran}

\usepackage[utf8]{inputenc}
\usepackage{amsmath}
\usepackage{graphicx}
\usepackage{hyperref}
\usepackage{listings}
\usepackage{booktabs}
\usepackage{morewrites} % For additional write streams
\usepackage{xcolor} % For colored text
\usepackage{geometry} % For margin adjustments
\usepackage{tabularx} % Advanced tables
\usepackage{comment} % Allows multi-line comments

\reserveinserts{1000} % Increase inserts for large documents

\begin{document}
\title{Porting an LLM based Application from ChatGPT to an On-Premise Environment}

\author{
\IEEEauthorblockN{Teemu Paloniemi}
\IEEEauthorblockA{%\textit{Faculty of Information Technology} \\
\textit{University of Jyväskylä}\\
Jyväskylä, Finland \\
teemu.a.j.paloniemi@student.jyu.fi}
\and
\IEEEauthorblockN{Manu Setälä}
\IEEEauthorblockA{
\textit{Solita}\\
Tampere, Finland \\
manu.setala@solita.fi}
\and
\IEEEauthorblockN{Tommi Mikkonen}
\IEEEauthorblockA{%\textit{Faculty of Information Technology} \\
\textit{University of Jyväskylä}\\
Jyväskylä, Finland \\
tommi.j.mikkonen@jyu.fi}

}

\maketitle

\begin{abstract}
Given the data-intensive nature of Machine Learning (ML) systems in general, and Large Language Models (LLM) in particular, using them in cloud based environments can become a challenge due to legislation related to privacy and security of data. Taking such aspects into consideration implies porting the LLMs to an on-premise environment, where privacy and security can be controlled. In this paper, we study this porting process of a real-life application using ChatGPT, which runs in a public cloud, to an on-premise environment. The application being ported is AIPA, a system that leverages Large Language Models (LLMs) and sophisticated data analytics to enhance the assessment of procurement call bids.
%and propose a framework for the porting process that helps in building suitable system fitting to one's business needs. 
The main considerations in the porting process include transparency of open source models and cost of hardware, which are central design choices of the on-premise environment. In addition to presenting the porting process, we evaluate downsides and benefits associated with porting. 
\end{abstract}

% Peer review title
\IEEEpeerreviewmaketitle

\section{Introduction}

Machine Learning (ML) and Artificial Intelligence (AI) have become widely used techniques in various applications. Given the data-intensive nature of such systems, their design, development, and operation processes must also consider data and cybersecurity related IPR and legislation. In the EU, relevant laws include the Data Governance Act (DGA) and the Data Act, as well as cybersecurity directives like the NIS2 Directive (NIS2) and the Cybersecurity Act (CSA). Furthermore, the EU AI Act oversees the responsible use of AI, ML, and related technologies in general in the EU context. 

To meet the above requirements, the developers -- especially those dealing with high-risk, business critical AI systems -- must conform to assessments and post-market monitoring. Ensuring human intervention capability in AI system design is crucial for decision-making when necessary. Hence, legal and ethical frameworks directly influence system design, as some public clouds should not process certain private data in any form \cite{ghorbel2017privacy}. Furthermore, it has also been pointed out that IoT systems in general lend themselves to considerations with respect to privacy, compliance, and ethics in their design \cite{pyry2024icwe}.

Large Language Models (LLM) are systems utilizing ML/AI to understand and generate human language \cite{chang2023survey}. Trained on vast amounts of text data, LLMs are able to perform a wide range of language-related tasks, such as answering questions and writing essays, for example. LLMs are capable of understanding the nuances of language, enabling them to  assist in various domains like education, business, and research.

In this paper, we study how a system created with ChatGPT, probably the best known LLM service at the moment, can be ported to an on-premise environment. By doing so, the motivation is to mitigate associated legal requirements in a given industrial use case. The goal is to understand the necessary technical steps and the associated losses in accuracy when an LLM system is no longer used from a public cloud like ChatGPT, but an on-premise version of the corresponding model is created and deployed instead. 
%Doing so enables data-intensive applications also in use cases when there are concerns about sharing the data in public clouds, for privacy or  cost related reasons, for instance. 
As an example application, we use 
an Artificial Intelligence Procurement Assistant (AIPA), a system that matches company profiles with those bids that are the most interesting and relevant. In doing so, the system leverages LLMs and sophisticated data analytics to process the assessment of public procurement call bids and other public funding opportunities \cite{waseem2023artificial}.

%The rest of the paper is structured as follows. In Section 2, we provide the background and the motivation for the paper. In Section 3, we introduce our research approach, including research context and questions, and applied research methods. In Section 4, we provide a technical insight to porting an LLM from cloud to an on-premise environment. In Section 5, we give an extended discussion on lessons learned in the process. Finally, in Section 6, we draw some final conclusions.

\section{Background and Motivation}

%The background of this work is divided to three parts. First, we discuss the relation of public and on-premise cloud services, their strengths, weaknesses, and limitations. Then, we give an overview of Large Language Models (LLM), which have become a common technology for various types of applications. Finally, we discuss porting in the context of ML, where data and other infrastructure play an important role.

\subsection{Large Language Models}

Large Language Models (LLM) are ML models trained to understand, generate, and interact with human language in a meaningful way \cite{chang2023survey}. These models are based on deep learning techniques \cite{lecun2015deep}, particularly a subset known as neural networks \cite{dongare2012introduction}, which allow them to process and learn from vast amounts of textual data.  The evolution of LLMs has marked a significant milestone in natural language processing (NLP), transforming how machines interact with human language \cite{chang2023survey}.

LLMs are trained on huge datasets, where text from books, articles, websites, and other forms are used as the training material \cite{hadi2024large}.  By analyzing this data, the model develops an understanding of not just grammar and syntax but also the contextual meaning behind words and phrases \cite{chang2023survey}. Therefore, LLMs perform tasks such as generating coherent essays, solving problems based on given prompts, or even engaging in conversations that appear remarkably natural and human-like. 

%As LLMs have advanced, they have become increasingly adept at specialized tasks beyond basic text generation, such as technical writing and customer service automation \cite{hadi2024large}. Models like GPT-3.5 and BERT have advanced the natural language processing, allowing machines to understand and create text that is similar to human \cite{mohamadi2023chatgpt}. These models have demonstrated their  their effectiveness in various tasks, including translating languages, generating text, answering questions, and analyzing sentiment \cite{alberts2023large}. Therefore, companies have started to use LLMs to analyze customer data, generate personalized content, and automate repetitive language-based tasks. This ability to adapt to various domains has made LLMs important in various fields. 

Despite their impressive capabilities, LLMs do face limitations and challenges \cite{hadi2024large}. They can sometimes generate biased or incorrect information since they are trained on data that may contain inaccuracies or societal biases. Moreover, LLMs include traces of training data, which means that their deployment requires understanding the restrictions related to privacy and regulatory concerns \cite{yao2024survey}.

\subsection{Public versus On-Premise Cloud Services}

Modern public cloud services are efficient, scalable, and straightforward to deploy and use. However, they may not be feasible or desirable for some use cases, especially when considering privacy and confidentiality requirements.  For example, the EU's General Data Protection Regulation (GDPR) defines personal data and outlines the obligations of companies when processing it. Furthermore, EU's GAIA-X \cite{braud2021road} and IDSA \cite{IDSA2024,otto2020creating} introduce numerous principles how to accomplish private and trusted data spaces, with interoperability that can be customized. Finally, integration At higher levels of classified information, such as governmental secrets, handling requirements -- like Katakri in Finland \cite{katakri} -- mandate that data remain within controlled spaces and not be moved elsewhere. %\color{red}  [https://um.fi/documents/35732/0/FINAL+-+Katakri-2020\_201218\_en.pdf/705d2bc6-6f1b-90dd-52e1-1ef97dae0623] \color{black}  
Finally, when considering IoT systems -- in particular those that feature edge intelligence \cite{peltonen2022many} -- it is possible that certain configurations require local data processing instead of central cloud. 

%In addition, there are other concerns, which are not important from the regulatory perspective, but have an impact on running the service, response times, energy consumption, and so on. These require careful orchestration of functions in the system, especially in the context of IoT systems \cite{peltonen2022many}. Therefore, the existence of such variety of potentially conflicting, yet evolving requirements makes the use of cloud computing extremely difficult in numerous real-life use cases. 

An on-premise environment gives the organization running the system full control over security and data location. However, such approach also implies that the organization takes responsibility over everything else related to running the on-premise environment. This means facilities that are given by design by a public cloud need to be implemented in the on-premise context. This can be a challenging task, as considerations regarding the design choices for the environment and software used are critical. Moreover, in addition to the technical implementation as such, there are also considerations about the accuracy and reliability of the results when porting an LLM system from ChatGPT to an on-premise environment.  

\subsection{Porting ML Models}

Porting software means transferring a piece of code to a different environment \cite{lecarme1986software}. Unlike traditional software, which can often be ported with relative ease, machine learning (ML) systems present unique challenges when adapting to new contexts, as already hinted in  \cite{mikkonen2021machine}. By their nature, these systems are tightly coupled with  the specific data, infrastructure, and hardware they were originally trained on. 

When porting an ML model to a different environment, such as moving from a cloud-based platform to an on-premise setup, the model's performance can be affected by differences in computational power, data pipelines, and system architecture, among other things \cite{kotilainen2023towards}. Retraining or fine-tuning is often necessary to adjust the model to the new environment, ensuring that it can still deliver accurate and reliable results. Finally, changes needed for porting can be of varying complexity. For instance, porting the same underlying model the same like in \cite{nasari2023porting,fassold2024porting} and training a totally new model that replaces another one require different activities.

%Depending on system complexity, the above adaptation can be both time-consuming and resource-intensive \cite{mungoli2023scalable}. Retraining the model might require access to the original data, additional computational resources, and specialized knowledge in model optimization. Even after retraining, the model's behavior could change due to differences in data distribution, hardware performance, or network configurations. Engineers must monitor the model closely to detect any degradation in accuracy or performance and make further adjustments as needed. Hence, moving an ML system to a new environment may not be a simple task.

%and with a suitably steady load, the costs will outweigh on-premises solutions over the long term. If a company already has existing on-premises systems and infrastructure, it may be easier, architecturally and in terms of data transfer volumes, to integrate an on-premises AI solution into existing environments. 

%The growth of AI computing power has been so rapid that within a few years it might be possible to use language models even in edge computing devices. The knowledge generated by this study can then be used in edge solutions where it is not possible to have a network connection.

\section{Research Approach}

%In this section, we first provide an overview of the research context. Then, we define research questions for this work. Finally, we discuss research methods that have been applied in this work.

\subsection{Research Context}

This paper studies porting of an LLM system from ChatGPT, which runs in a public cloud, to an on-premise environment. As an example, we use Artificial Intelligence Procurement Assistant (AIPA) \cite{waseem2023artificial}, 
a system that crawls company web pages and EU wide public tenders to find matches between them\footnote{\href{https://ted.europa.eu/en/}{https://ted.europa.eu/en/}}. 
%Figure \ref{fig:ui} shows the user interface of the system, displaying today's most matching tenders for a company profile.

\begin{comment}
\begin{figure*}[t] 
    \centering 
    \includegraphics[width=2\columnwidth]{AIPA_results_page.png} 
%    \caption{AIPA in action, with interaction between different subsystems visualized.} 
    \caption{ACME user interface, with list of today's best tenders on the left, and an opened tender announcement on the right.} 
    \label{fig:ui}
%    \caption{AIPA high-level architecture and its end users.} 
\end{figure*}
\end{comment}

\begin{figure}[t] 
    \centering 
    \includegraphics[width=1.0\columnwidth]{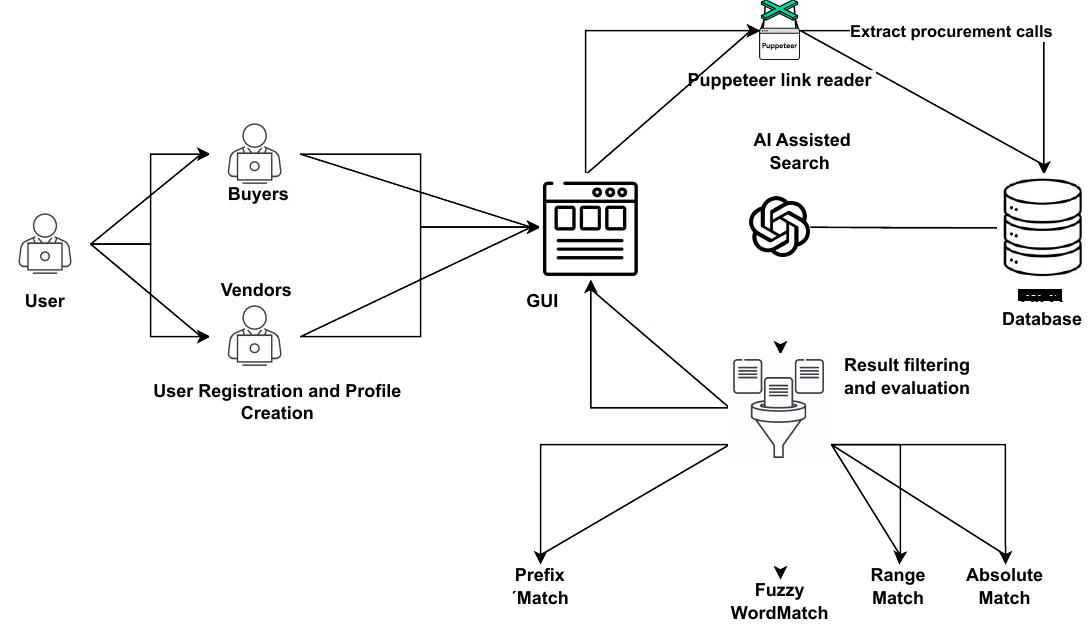} 
%    \caption{AIPA in action, with interaction between different subsystems visualized.} 
    \caption{AIPA high-level architecture. Running in the cloud, AI assisted search function is the key feature of the system.} 
    \label{fig:architecture}
%    \caption{AIPA high-level architecture and its end users.} 
\end{figure}

The high-level architecture of AIPA 
is presented in Figure \ref{fig:architecture}. The system utilizes ChatGPT to extract search parameters from company profiles,
%(Figure \ref{fig:profile}), 
with each piece of the profile consisting of free-form text. 
%(Figure \ref{fig:strategy}). 
These parameters are then employed to conduct searches from the AIPA database, which is constantly updated with procurement information from TED and other similar procurement websites. This optimization is crucial for efficiently searching through large volumes of documents, as loading everything from online on the need basis would introduce serious delays in search operations. The AI system that we have created comprises multiple GPT agents, with distinct roles and prompts to handle various tasks such as translation, keyword extraction, and generating similar words. These agents can be run as distributed tasks rather than monolithic ones, thus contributing to improved performance.

\begin{comment}
\begin{figure}[t] 
    \centering 
    \includegraphics[width=1.0\columnwidth]{AIPA_multiple_input.png} 
%    \caption{AIPA in action, with interaction between different subsystems visualized.} 
    \caption{ACME company profile formation, based on online resources.} 
    \label{fig:profile}
%    \caption{AIPA high-level architecture and its end users.} 
\end{figure}

\begin{figure}[t] 
    \centering 
    \includegraphics[width=1.0\columnwidth]{AIPA_extract_page_info.png} 
%    \caption{AIPA in action, with interaction between different subsystems visualized.} 
    \caption{Sample item in ACME company profile, given in natural language and parsed with LLMs.} 
    \label{fig:strategy}
%    \caption{AIPA high-level architecture and its end users.} 
\end{figure}
\end{comment}

%\subsection{Design goals}
The AIPA implementation, presented in \cite{waseem2023artificial}, was designed such that it fully relied on ChatGPT. However, upon discussions with a target company, three special needs were raised that motivated us to consider on-premise design. These were (i) enhanced security, or how to make sure that no other company is able to track down what a particular company is interested in; (ii) deeper customization, so that it would be possible to build company specific modifications; and (iii) resource efficiency of operating LLMs in an on-premise environment, instead of running everything with ChatGPT which turned out to be expensive.  

\subsection{Research Questions}

Based on the above background and research context, we are interested in answering to the following research questions:
\begin{itemize}
    \item[RQ1:] What steps are needed to port LLM related parts of 
    AIPA from ChatGPT to on-premise environment?
    \item[RQ2:] What design decisions are necessary in each step of the process?
    \item[RQ3:] How closely related results are produced by the original and the ported application? 
\end{itemize}

\subsection{Research Methods}

The research methods applied in this paper are case study and design science research. A case study \cite{runeson2009guidelines} is an in-depth analysis of am individual, event, or situation, often used in research to explore complex issues.  Case studies offer a focused examination of real-world contexts, making them valuable for understanding unique or rare cases, as well as drawing insights that might not be possible through broader research methods. In general, case studies are widely used to analyze real-life challenges and solutions. By examining specific examples, one can explore how theories apply in practical scenarios, what decision-making processes are involved, and how different factors interact in a particular situation. Therefore, case studies are a powerful tool for experiential research, where the goal is to probe the limits of existing design space, for instance.

Design Science Research (DSR) \cite{peffers2007design} is a methodology focused on solving complex problems through the creation and evaluation of innovative artifacts, such as models, frameworks, methods, or systems. The primary goal of DSR is to generate knowledge that improves the design and performance of these artifacts while addressing real-world problems. This methodology combines both scientific rigor and practical relevance by integrating theory-driven research with hands-on experimentation, with the resulting the artifact evaluated in real-world or simulated environments. % DSR emphasizes the generation of new knowledge, both in terms of theoretical contributions to the academic field and practical solutions that can be applied in industry. 

\section{Design and Implementation}

%This section first defines a porting strategy we decided to use for the case study. Then, it discusses the different phases of porting, including preparation for porting, actual implementation, and finally deployment and evaluation.

%in this paper we aim at investigating the portability of the generative AI cloud solution to an on-premises solution and also to determine what hardware requirements the on-premises solution must meet to provide a service that is fast enough to be used.  

%The following subsections intrd, the model and hardware selection, and an overview of the porting process.

%The software used in the study was derived from the \textit{Artificial Intelligence Procurement Assistant} (AIPA) application \cite{waseem2023artificial}, evolving into an AI assistant recommending European Union research funding instruments. This system worked as an example application of proven use case for cloud LLMs in software, thus making it a fitting platform to demonstrate the porting process. 

\subsection{Porting Strategy}

\begin{figure}[t] 
    \centering 
    \includegraphics[width=1.0\columnwidth]{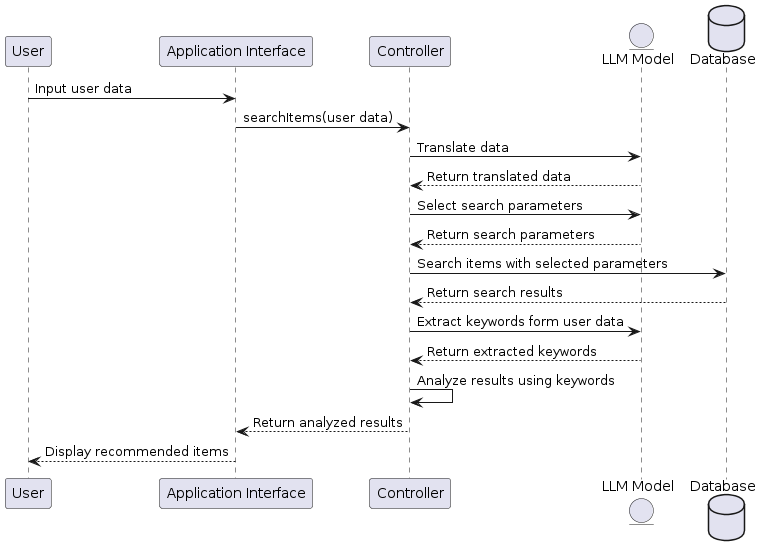} 
    \caption{AIPA in action, with subsystem interaction visualized.} 
    \label{fig:operation}
\end{figure}

AIPA, the system to be used in this case study, is a traditional web application. The system consists of a client that simply displays data to the user, and of a server that implements the business logic. The business logic includes conversation between the server controller, the database that holds the information on public tenders, and cloud LLM application programming interface (API). The original version of AIPA was build around ChatGPT, which runs as a public service.
Figure \ref{fig:operation} illustrates the operational flow of the application between these components. The core of the business logic of the application lies in the communication with the LLM API -- as shown in the figure, six distinct requests and responses are send over the Internet to the cloud API and back.

%This communication exposes the program to potential data leaks and might transfer and store user data to the cloud servers.

%\subsection{Porting}

\begin{figure}[t] 
    \centering 
    \includegraphics[width=1.0\columnwidth]{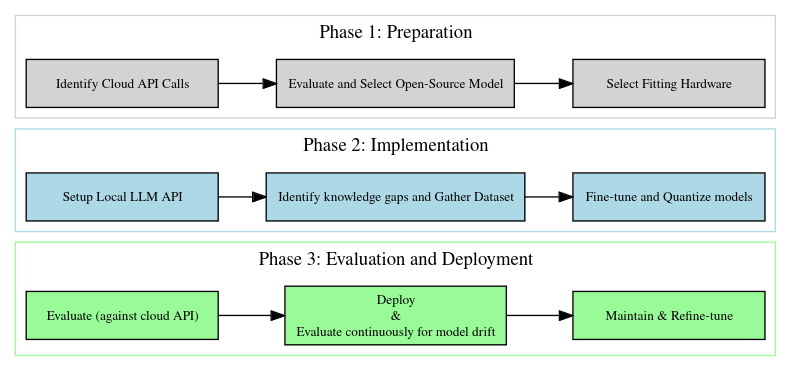} 
    \caption{Steps of the porting process.}  
    \label{fig:porting}
\end{figure}

Several strategies were considered for porting the LLM part of AIPA. 
The alternatives were analyzed, leading to the identification of three main steps necessary for porting the system from ChatGPT to an on-premise environment. As visualized in Figure \ref{fig:porting}, these steps include: preparation, or refactoring the baseline system to simplify porting and selecting components for the on-premise environment; implementation, or selecting and designing the necessary new subsystems in the on-premise environment; and deployment to the new environment followed by an associated evaluation. 

\subsection{Preparation}

The preparation step includes three activities. These are (i) code changes; (ii) hardware specification; and (iii) model selection, discussed in more detail in the following.

\subsubsection{Code changes} At the beginning of the preparation step, we refactored the system by wrapping the cloud API with a local subroutine. Then, we traced all the code lines that made calls to the cloud API, and replaced direct API calls with calls to the wrapper, thus localizing cloud APIs for modifications when needed. 

\subsubsection{Hardware Specifications} %Previous studies have demonstrated the resource efficiency of cloud solution in ML and other data science applications \cite{chalker2020cloud} \cite{juhasz2021quantitative}. Clouds are often designed to serve the needs of large scalable solutions. The benefit of cloud hardware therefore diminishes as we get closer to small teams and individual actors. 
Machine learning models have myriad architectures and their required hardware varies considerably. For example neural network based models such as the LLMs require highly parallel computing when other methods like support vector machines or regression models can sufficiently be deployed on smaller compute. Therefore, when selecting hardware for ML purposes, we had tp take into account the model we plan to use, the business logic of the designed system, and resources we have available.

ML applications in low-compute environments have been studied and for example \cite{kaiser2021towards} describes methodically the use of ML in small and medium-sized enterprises (SMEs) in a cost effective way. The main idea being that simple solutions and off the shelf equipment can be adequate for many business use cases. We build on top of this work by considering consumer graphics processing units (GPUs) as affordable business hardware as they are designed to be installed and used by layperson, they are effective with current LLM architectures and are reasonably priced compared to high-end data center GPUs or application specific integrated circuits (ASICs). In our system, the models were used primarily for inference purposes, so massive datasets and extensive training compute were not necessary. Additionally, as the system was intended for use by small teams or individuals, there was no need for a large scale compute cluster.

\subsubsection{Model Selection} Open-source ML models come in various shapes and sizes. When porting the AIPA system, the design decisions, including model selection, were guided by the current regulation and application logic. Hence, selecting a suitable ML model required collecting model metadata, such as the training dataset, model architecture, and licence, and comparing them against system requirements and legal constraints. For AIPA, we decided that we wish to minimize the probability of our system generating private or sensitive information, which could be present in the training dataset. GDPR holds the controller of the system accountable for compliance, so this is a realistic requirement for several real-life systems. 

Transparency for model metadata has been advocated in studies like \cite{mitchell2019model} and \cite{crisan2022interactive}. Regulations that follow this work can be found in the EU AI act parts (89), (102), and (103) that also highlight transparency and openness of open-source models \cite{edwards2021eu}. We found that the model cards and current legislation provide valuable information on the model creation but lack any mention on model behaviour, as also pointed out in \cite{pyry2024icwe}. Therefore, even if the model training process is transparent, our system must take into account any security anomalies formed as a product of using the model.   

In addition, model architecture and especially its size are important factors for the selection process, as the available computing capacity is often limited or the business logic favors speed over quality. Smaller models intuitively require less computation, but might not carry the same accuracy as bigger models do. 

A comprehensive way to evaluate model suitability is to use general benchmark frameworks designed to differentiate models from each other in various domain-specific tasks. These frameworks can combine several aspects of model metadata and performance in an aggregate format, thereby aiding the selection process. %Moreover, a key factor that has an effect on model selection is formed by security and ethics aspects of using an open-source on-premises ML models, which echoes the limitations related to computing capacity. 

We used Eleuther AI Language Model Evaluation Harness framework \cite{eval-harness} and HuggingFace Open LLM Leaderboard \cite{open-llm-leaderboard} as model qualifiers. With the above concerns in mind, we selected the most performant small and large models to test them in our system. Both models were either created inside EU or their training datasets were open-sourced in a transparent fashion. 

\subsection{Implementation}

Once the activities in the preparation step were completed, we next proceeded to implementation step. In this step the aim was namely implement the prepared system, test it and identify any gaps in model performance in order to fix them via customization. 

First we implemented a local API that matches the cloud API in format and functionality. Various libraries exist for this purpose \cite{kwon2023efficient,llama.cpp,Aminabadi2022deepspeed}. We used \verb|llama.cpp| \cite{llama.cpp}. The library enables LLM inference with minimal setup and state-of-the-art performance on a wide variety of hardware both locally and in the cloud, and was selected for its simplicity. In addition, support for different compute requirements was a factor; \verb|llama.cpp| is designed to work with CPUs, and allows for GPU acceleration. This way the resulting program could be tested in multiple scenarios, such as individual laptops, workstations or larger compute clusters. 

As mentioned, the system had to take into account any security anomalies in the generated data. We therefore used the models only in communication between the system server-side and the database. Therefore, all the information used in the client side came from the curated database, not directly from the language model, thus minimizing the risk for data privacy violations.

Usage testing was done by running the small model as part of the system to check for any gaps in information flow from the user to the model and back to the system. We found that smaller models are useful in testing the system as they require less compute and result in faster generation speeds therefore enabling faster iteration on development. We did not identify any security issues during our tests, as no outward communication left the system and no messages were saved. The biggest problems arose with model hallucination on outdated or incorrect information on our domain of interest. Based on this, we concluded that the training data did not contain recent updates on that specific field and had a low probability of working as part of the system. Therefore the model needed fine-tuning. 

We gathered a dataset of various documents from our database. The dataset was formed by feeding bits of the documents to LLM that was instructed to generate meaningful questions which had answers in the text. The questions and answers were then paired to form data points we could use for the fine-tuning. This dataset was then used to fine-tune the smaller model with the larger compute using PyTorch \cite{pytorch} and Low-Rank Adaptation (LoRA) technique \cite{hu2021lora}. When tested again we found that this increased the relevancy of the results. We had no available resources to fine-tune the larger model and hence found that smaller models are better for applications with constantly changing information baseline as they are also faster to customize. 
%Through this process, we ensured some level of control over the model behaviour.

\subsection{Deployment and Evaluation}

In the deployment and evaluation step we backtracked all the steps of the porting process and evaluated the achieved level of performance and the design goals. As we identified all the outgoing requests from the original software and selected a model according to our best knowledge on current legislation the system achieves the security of a typical on-premise software. Hardware and model were also aligned with the business logic minimizing the negative effect of open-source models on accuracy. However, this was only tested with human users, not systematically compared with some underlying benchmark. Some private or sensitive data leaks to the client side could not be ruled out in the model generation. To compensate this possibility, we made a design decision that the model is only used on the server side of the system.  

\section{Discussion}

In summary, the system achieved all the design goals that were defined at the beginning of the porting process. Based on the above porting, we next address the lessons we have learned in the process. Then, we discuss the essential limitations of this case study and possible directions for future research.

%\subsection{Lessons learned}

%    - the results of your research,

%    - a discussion of related research
%This study followed the work of ML/AI tooling research \cite{waseem2023artificial} \cite{mitchell2019model} \cite{kaiser2021towards} by considering consumer grade hardware and open-source models. The case was motivated by the current data and AI legislation and businesses need for more secure and controllable systems \cite{ghorbel2017privacy} \cite{kotilainen2023webassembly}. 
    
%    - How can future research build on these observations? What are the key experiments that must be done? 

The key lesson we have learned in this case study is that porting LLMs from ChatGPT to on-premise environment is feasible. Moreover, the steps that are needed in the process are not very complex, and many of them resemble the traditional porting process, with additional APIs introduced to support the process. Our biggest concern was associated with model training, but this turned out to be easier than expected, due to wide availability of existing model and tooling options. While the resulting system is not as powerful as the original version, we expect that with more training or larger model, this can be overcome. Moreover, one can argue if the systems should give exactly the same answers, due to the stochastic nature of many ML systems.

The on-premise version has also been taken to use in our industry partner's operations. This would not have been possible with the ChatGPT based version. This was mainly due to associated cost issues, although there were also some privacy and security related issues involved. We expect that with increasing understanding on LLMs non-functional properties, including in particular privacy and costs mentioned here, there will be APIs that enable replacement of one LLM system with another one, with minimal changes in code. Experimenting this with different architectural approaches is left as future work.

Finally, it is a key observation that dealing with non-functional properties of ML models in general is becoming a part of software engineering practice. Software engineers need to grasp what are the relevant restriction for using a certain kind of an ML model and its embedding in the software architecture, in collaboration with data scientists that deal with the models themselves.

\section{Conclusions}

Modern ML systems, such as LLMs, can include traces of material they have been trained on. Therefore, it is not self-evident when they can be used in a public cloud, and when an on-premise environment is a better fit for the system. Unlike with classical software, porting the intelligent part of the application to a new context can be a laborous task. 
%    - the results of your research,
By porting a ChatGPT based system to an on-premise environment, we showed that an on-premise environment can be used to provide more secure and customized solution compared to using cloud based proprietary ML models. In addition, costs are also a major issue that can have an impact on the selection between a public cloud and an on-premise environment.

%The security emerges from two factors. Firstly, the system was removed from the open internet and therefore mitigated the risk for leaking sensitive and private information out of the system. Secondly, the system could be build considering the legislation and for example in model selection the transparency of open-source models help designing a system detailed to a case specific requirements. 

% Bibliography
\bibliographystyle{ieeetr}
\bibliography{references}

\begin{thebibliography}{10}

\bibitem{ghorbel2017privacy}
A.~Ghorbel, M.~Ghorbel, and M.~Jmaiel, ``Privacy in cloud computing environments: a survey and research challenges,'' {\em The Journal of Supercomputing}, vol.~73, no.~6, pp.~2763--2800, 2017.

\bibitem{pyry2024icwe}
P.~Kotilainen, A.~Mehraj, T.~Mikkonen, and N.~Mäkitalo, ``The programmable world and its emerging privacy nightmare,'' in {\em International Conference on Web Engineering}, Springer, 2024.

\bibitem{chang2023survey}
Y.~Chang, X.~Wang, J.~Wang, Y.~Wu, L.~Yang, K.~Zhu, H.~Chen, X.~Yi, C.~Wang, Y.~Wang, {\em et~al.}, ``A survey on evaluation of large language models,'' {\em ACM Transactions on Intelligent Systems and Technology}, 2023.

\bibitem{waseem2023artificial}
M.~Waseem, T.~Das, T.~Paloniemi, M.~Koivisto, E.~R{\"a}s{\"a}nen, M.~Set{\"a}l{\"a}, and T.~Mikkonen, ``Artificial intelligence procurement assistant: Enhancing bid evaluation,'' in {\em International Conference on Software Business}, pp.~108--114, Springer Nature Switzerland Cham, 2023.

\bibitem{lecun2015deep}
Y.~LeCun, Y.~Bengio, and G.~Hinton, ``Deep learning,'' {\em nature}, vol.~521, no.~7553, pp.~436--444, 2015.

\bibitem{dongare2012introduction}
A.~Dongare, R.~Kharde, A.~D. Kachare, {\em et~al.}, ``Introduction to artificial neural network,'' {\em International Journal of Engineering and Innovative Technology (IJEIT)}, vol.~2, no.~1, pp.~189--194, 2012.

\bibitem{hadi2024large}
M.~U. Hadi, Q.~Al~Tashi, A.~Shah, R.~Qureshi, A.~Muneer, M.~Irfan, A.~Zafar, M.~B. Shaikh, N.~Akhtar, J.~Wu, {\em et~al.}, ``Large language models: a comprehensive survey of its applications, challenges, limitations, and future prospects,'' {\em Authorea Preprints}, 2024.

\bibitem{yao2024survey}
Y.~Yao, J.~Duan, K.~Xu, Y.~Cai, Z.~Sun, and Y.~Zhang, ``A survey on large language model {(LLM} security and privacy: The good, the bad, and the ugly,'' {\em High-Confidence Computing}, p.~100211, 2024.

\bibitem{braud2021road}
A.~Braud, G.~Fromentoux, B.~Radier, and O.~Le~Grand, ``The road to european digital sovereignty with gaia-x and idsa,'' {\em IEEE network}, vol.~35, no.~2, pp.~4--5, 2021.

\bibitem{IDSA2024}
{International Data Spaces Association}, ``Home -- international data spaces.'' \url{https://internationaldataspaces.org/}, 2024.
\newblock retrieved 2024-09-16.

\bibitem{otto2020creating}
B.~Otto, ``Creating data spaces based on gaia-x and ids,'' {\em Hitachi Research Institute Journal}, vol.~15, no.~2, pp.~32--37, 2020.

\bibitem{katakri}
{National Security Authority of Finland}, ``{Katakri: Information Security Audit Tool for Authorities}.'' {Available at https://um.fi/information-security-auditing-tool-for-authorities-katakri}, 2020.

\bibitem{peltonen2022many}
E.~Peltonen, I.~Ahmad, A.~Aral, M.~Capobianco, A.~Y. Ding, F.~Gil-Castineira, E.~Gilman, E.~Harjula, M.~Jurmu, T.~Karvonen, {\em et~al.}, ``The many faces of edge intelligence,'' {\em IEEE Access}, vol.~10, pp.~104769--104782, 2022.

\bibitem{lecarme1986software}
O.~Lecarme and M.~Pellissier~Gart, {\em Software portability}.
\newblock McGraw-Hill, Inc., 1986.

\bibitem{mikkonen2021machine}
T.~Mikkonen, J.~K. Nurminen, M.~Raatikainen, I.~Fronza, N.~M{\"a}kitalo, and T.~M{\"a}nnist{\"o}, ``Is machine learning software just software: A maintainability view,'' in {\em Software Quality: Future Perspectives on Software Engineering Quality: 13th International Conference, SWQD 2021, Vienna, Austria, January 19--21, 2021, Proceedings 13}, pp.~94--105, Springer, 2021.

\bibitem{kotilainen2023towards}
P.~Kotilainen, V.~Heikkil{\"a}, K.~Syst{\"a}, and T.~Mikkonen, ``Towards liquid ai in iot with webassembly: a prototype implementation,'' in {\em International Conference on Mobile Web and Intelligent Information Systems}, pp.~129--141, Springer, 2023.

\bibitem{nasari2023porting}
A.~Nasari, L.~Zhai, Z.~He, H.~Le, S.~Cui, D.~Chakravorty, J.~Tao, and H.~Liu, ``Porting ai/ml models to intelligence processing units (ipus),'' in {\em Practice and Experience in Advanced Research Computing}, pp.~231--236, 2023.

\bibitem{fassold2024porting}
H.~Fassold, ``Porting large language models to mobile devices for question answering,'' {\em arXiv preprint arXiv:2404.15851}, 2024.

\bibitem{runeson2009guidelines}
P.~Runeson and M.~H{\"o}st, ``Guidelines for conducting and reporting case study research in software engineering,'' {\em Empirical software engineering}, vol.~14, pp.~131--164, 2009.

\bibitem{peffers2007design}
K.~Peffers, T.~Tuunanen, M.~A. Rothenberger, and S.~Chatterjee, ``A design science research methodology for information systems research,'' {\em Journal of management information systems}, vol.~24, no.~3, pp.~45--77, 2007.

\bibitem{kaiser2021towards}
J.~Kaiser, G.~Terrazas, D.~McFarlane, and L.~de~Silva, ``Towards low-cost machine learning solutions for manufacturing smes,'' {\em AI \& society}, pp.~1--7, 2021.

\bibitem{mitchell2019model}
M.~Mitchell, S.~Wu, A.~Zaldivar, P.~Barnes, L.~Vasserman, B.~Hutchinson, E.~Spitzer, I.~D. Raji, and T.~Gebru, ``Model cards for model reporting,'' in {\em Proceedings of the conference on fairness, accountability, and transparency}, pp.~220--229, 2019.

\bibitem{crisan2022interactive}
A.~Crisan, M.~Drouhard, J.~Vig, and N.~Rajani, ``Interactive model cards: A human-centered approach to model documentation,'' in {\em Proceedings of the 2022 ACM Conference on Fairness, Accountability, and Transparency}, pp.~427--439, 2022.

\bibitem{edwards2021eu}
L.~Edwards, ``The {EU AI Act}: a summary of its significance and scope,'' {\em Artificial Intelligence (the EU AI Act)}, vol.~1, 2021.

\bibitem{eval-harness}
L.~Gao, J.~Tow, B.~Abbasi, S.~Biderman, S.~Black, A.~DiPofi, C.~Foster, L.~Golding, J.~Hsu, A.~Le~Noac'h, H.~Li, K.~McDonell, N.~Muennighoff, C.~Ociepa, J.~Phang, L.~Reynolds, H.~Schoelkopf, A.~Skowron, L.~Sutawika, E.~Tang, A.~Thite, B.~Wang, K.~Wang, and A.~Zou, ``A framework for few-shot language model evaluation,'' 12 2023.

\bibitem{open-llm-leaderboard}
E.~Beeching, C.~Fourrier, N.~Habib, S.~Han, N.~Lambert, N.~Rajani, O.~Sanseviero, L.~Tunstall, and T.~Wolf, ``Open {LLM} leaderboard.'' \url{https://huggingface.co/spaces/HuggingFaceH4/open_llm_leaderboard}, 2023.

\bibitem{kwon2023efficient}
W.~Kwon, Z.~Li, S.~Zhuang, Y.~Sheng, L.~Zheng, C.~H. Yu, J.~E. Gonzalez, H.~Zhang, and I.~Stoica, ``Efficient memory management for large language model serving with pagedattention,'' 2023.

\bibitem{llama.cpp}
G.~Georgi {\em et~al.}, ``Llama.cpp,'' 2023.

\bibitem{Aminabadi2022deepspeed}
R.~Y. Aminabadi, S.~Rajbhandari, A.~A. Awan, C.~Li, D.~Li, E.~Zheng, O.~Ruwase, S.~Smith, M.~Zhang, J.~Rasley, and Y.~He, ``Deepspeed-inference: Enabling efficient inference of transformer models at unprecedented scale,'' in {\em SC22: International Conference for High Performance Computing, Networking, Storage and Analysis}, pp.~1--15, 2022.

\bibitem{pytorch}
{{The PyTorch Project}}, ``{{PyTorch web site}},'' 2024.
\newblock retrieved 2024-6-13.

\bibitem{hu2021lora}
E.~J. Hu, Y.~Shen, P.~Wallis, Z.~Allen-Zhu, Y.~Li, S.~Wang, L.~Wang, and W.~Chen, ``Lora: Low-rank adaptation of large language models,'' {\em arXiv preprint arXiv:2106.09685}, 2021.

\end{thebibliography}

\end{document}